\documentclass[twocolumn,aps,epsfig,nofootinbib]{revtex4}

\usepackage{graphicx}
\usepackage{epstopdf}
\usepackage{latexsym}
\usepackage{amssymb}
\usepackage{amsmath}
\usepackage{color}

\usepackage[center]{subfigure}

\begin{document}

 \newcommand{\bq}{\begin{equation}}
 \newcommand{\eq}{\end{equation}}
 \newcommand{\bqn}{\begin{eqnarray}}
 \newcommand{\eqn}{\end{eqnarray}}
 \newcommand{\nb}{\nonumber}
 \newcommand{\lb}{\label}
\newcommand{\PRL}{Phys. Rev. Lett.}
\newcommand{\PL}{Phys. Lett.}
\newcommand{\PR}{Phys. Rev.}
\newcommand{\CQG}{Class. Quantum Grav.}

\title{Gravitational quantum  effects in  light of BICEP2 results}

 \author{Tao Zhu}
\email{tao_zhu@baylor.edu}

\author{Anzhong Wang\footnote{Corresponding author}}
\email{anzhong_wang@baylor.edu}

 \affiliation{GCAP-CASPER, Physics Department, Baylor University, Waco, TX 76798-7316, USA\\
  Institute for Advanced Physics $\&$ Mathematics, Zhejiang University of Technology, Hangzhou, 310032, China}

\date{\today}


\begin{abstract}

Recently BICEP2 found  that the vanishing of the tensor-to-scalar ratio $r$ is excluded at $7\sigma$ level, and  its most likely value
is  $r=0.2^{+0.07}_{-0.05}$ at $1 \sigma$ level. This immediately causes a tension with the Planck constraint $r < 0.11$. In addition, 
it also implies that the inflaton (in  single field slow-roll inflation) experienced a Planck excursion during inflation  $\Delta\phi /M_{Pl} 
\geq {\cal{O}}(1)$, whereby the effective theory of inflation becomes questionable. In this brief report, we show that the inflationary 
paradigm is still robust, even after the quantum effects are taken into account. Moreover, these effects also help to relax the tension 
on the different values of $r$ given by BICEP2  and Planck.

\end{abstract}


\pacs{98.80.Bp; 98.80.Qc; 98.80.Cq; 04.50.Kd}

\maketitle

\section{Introduction}

The inflationary scenario provides a framework for solving the problems of the standard big bang cosmology and, most importantly, provides a causal mechanism for generating structure in the universe and the spectrum of cosmic microwave background (CMB) anisotropies.  In this picture, primordial density and gravitational wave fluctuations are created from quantum fluctuations during the inflation process. The former, which generates observational effects in temperature anisotropies in CMB, has been conformed by observations with spectacular precision  \cite{WMAP,PLANCK}. For the primordial gravitational wave fluctuations, which generates the unique primordial B-mode polarization of CMB, has now been observed by the ground based BICEP2 experiment  \cite{BICEP2}. The results of the experiment shows for the first time a non-zero value of the tensor-to-scalar ratio $r$ at $7\sigma$ C.L., and  estimated  
\bq
\lb{rvalue}
r=0.2^{+0.07}_{-0.05},
\eq
at $1 \sigma$ C.L.
 If it is confirmed, it will represent one of the most important discoveries in the field, and provides us a new diagnostic tools to probe the models of early universe, especially the models in the framework of quantum gravity.  

However, the BICEP2 results are   in tension with  the Planck constraint $r<0.11$  \cite{PLANCK}. As pointed out by BICEP2 group \cite{BICEP2}, this tension may be alleviated if a large negative running of scale spectral index is allowed. However, such a large negative running can not be produced in the framework of single field slow-roll inflation models. As the single-field slow-roll inflation has already been shown to be favored by the Planck data, it is very desirable to find  some mechanisms to relieve this tension. 

More important, a large $r$, such that given by Eq.(\ref{rvalue}),  also implies that  the effective field theory of inflation is problematic. This  is based on the analysis of the Lyth bound  \cite{lyth}, which states that in the slow-roll approximations, the change of the inflationary field $\phi$ during inflation is given by
\bqn
\lb{Lyth}
\frac{\Delta\phi}{M_{\text{pl}}} \simeq \sqrt{\frac{r}{8}} \Delta{N},   
\eqn
where  $M_{\text{pl}}$ denotes the Planck mass, and  $\Delta{N}$  the number of e-folds. If $r$ does not change as a function of $N$, this directly leads to \cite{lyth}
\bqn
\frac{\Delta \phi}{M_{\text{pl}}} \simeq \mathcal{O}(1) \sqrt{\frac{r}{0.01}}.
\eqn
Obviously,  $\Delta \phi$ exceeds the Planck scale if $r$ is of the order of Eq.(\ref{rvalue}).  As a result, the effective field theory of inflation with a potential $V(\phi)$, which consists of a derivative expansion of operators suppressed by Planck scale, becomes questionable.

In this brief report, we  take the point of view that $r$ is indeed large and of the order of Eq.(\ref{rvalue}), so that  the gravitational quantum effects   are important and necessary to be taken into account during the epoch of inflation. Once these effects are taken into account, we shall show that the inflation paradigm is still very robust, and almost scale-invariant can be easily produced. In addition, these effects   also help to relax the tension between the values of $r$ given, respectively,  by Planck and BICEP2. 

\section{Trans-Planckian Effects}

While quantum gravity has not been properly formulated, gravitational quantum  effects on inflation have been studied by various methods \cite{AA,BQ,Brandenberger2013CQG}. One of them is to assume that the dispersion relations  of scalar and tensor perturbations  during inflation become nonlinear \cite{Brandenberger1999,Martin2001,Niemeyer2001PRDR,Martin2004PRD}, similar to the study of Hawking radiation in black hole physics \cite{Unruh}. Remarkably, such relations can be realized naturally  in the framework of Horava-Lifshitz quantum gravity \cite{WM,KUY,Soda2009prl,Wang2010PRD,Huang2013,Zhu2013}, which is power-counting renormalizable by construction \cite{Horava}.

To show our above claim, let us first begin with the conventional linear dispersion relation $\omega_k^2(\eta)=k^2$, with which the inflationary perturbations $\mu_k(\eta)$ (scale or tensor) obey the equation
\bqn\lb{eom}
\mu_k''(\eta)+\left(\omega_k^2(\eta)-\frac{z''}{z}\right)\mu_k(\eta)=0,
\eqn
where a prime represents the differentiation with respect to conformal time $\eta$, $z(\eta)$ depends on the background and the types of perturbations, scalar or tensor, and usually is proportional to the cosmological scale factor $a(\eta)$. 

After quantum effects are taken into account,   the inflationary mode functions   still satisfy the above equation but with a nonlinear dispersion relation \cite{Brandenberger2013CQG,WM,KUY,Soda2009prl,Wang2010PRD,Huang2013,Zhu2013},  
\bqn\lb{omega}
\omega^2_k(\eta) &=& k^2 \left[1-\hat{b}_1 \left(\frac{k}{a M_*}\right)^2+\hat{b}_2 \left(\frac{k}{a M_*}\right)^4\right],
\eqn 
where $M_*$ is the relevant energy scale of trans-Planckian physics, $k$ is the comoving wavenumber of the mode, $\hat{b}_1$ and $\hat{b}_2$ are dimensionless constants. It should be noted that above nonlinear dispersion relation was initially applied to inflationary cosmology as a toy model  \cite{Brandenberger1999,Martin2001}, motivated from the studies of the dependence of black hole radiation on Planck-scale physics \cite{Unruh}. In particular, in the framework of Ho\v{r}ava-Lifshitz gravity, the above nonlinear dispersion can arise naturally, and the parameters $\hat b_1$ and $\hat b_2$ are expressed in terms of the coupling constants of the theory \cite{WM,KUY,Soda2009prl,Wang2010PRD,Huang2013,Zhu2013}. In order to make the theory power-counting renormalizable, one at least needs to include the sixth order spatial derivative terms ($\hat b_2$ term) in the dispersion relation (\ref{omega}), and in order to get a health UV limit, one requires $\hat{b}_2>0$.  
 
To proceed, it is more convenient to introduce the dimensionless variable $y=-k\eta$, and write  Eq.(\ref{eom}) in the form
 \cite{Olver1974,Olver1975}
\bqn\lb{eom58}
\frac{d^2\mu_{k}(y)}{dy^2}=\Big[g(y)+q(y)\Big]\mu_k(y),
\eqn
where the convergence of the solutions requires the following choice of the functions $g(y)$ and $q(y)$ \cite{Zhu},
\bqn
\lb{gy}
q(y)&=&-\frac{1}{4y^2}, \;\nb\\
g(y)&=&\frac{\nu^2}{y^2}-1+b_1\epsilon_*^2 y^2 -b_2\epsilon_*^4 y^4,
\eqn
where $\epsilon_*^2=H^2/M_*^2$, $H$ is the Hubble parameter.  Note that in the above we have set  $z''/z\equiv [\nu^2(\eta)-1/4]/\eta^2$ and $a \simeq - (1-\varepsilon+\mathcal{O}(\varepsilon^2))/(\eta H)$, here the slow roll parameter $\varepsilon$ has been absorbed in parameters $b_1\equiv \hat{b}_1 (1+2\varepsilon+\mathcal{O}(\varepsilon^2))$ and $b_2\equiv \hat{b}_2 (1+4 \varepsilon+\mathcal{O}(\varepsilon^2))$. 

To solve Eq.(\ref{eom58}),   we adopt the uniform asymptotic approximation developed recently in \cite{uniformPRL,Zhu}. 
We first note that  the function $g(y)$ could have three physically different cases, as illustrated   in Fig.1, depending on the number of  zeros or  turning points  in terms of the terminology given in  \cite{Olver1974}.
The corresponding roots  will be denoted by $y_0$, $y_1$ and $y_2$, where   $y_0$ is always real, while    $y_1$ and $y_2$ can be both real, or both complex, or coalesce into one double root. As have been shown in  \cite{Zhu}, the approximate solutions of mode functions depend on the properties of these turning points. 

\begin{figure}[t]
\centering
	{\includegraphics[width=75mm]{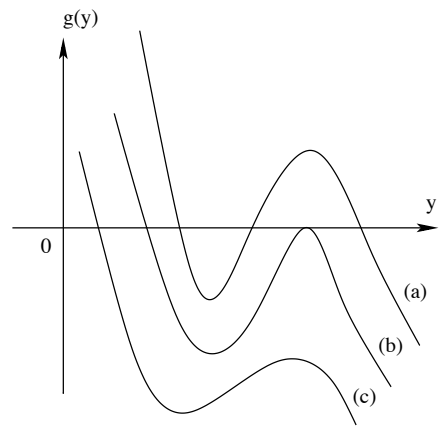}}
\caption{The function $g(y)$ defined by Eq. (\ref{gy}).
(a)  Three  different  single real  roots of the equation $g(y) = 0$.
(b) One single and one double real root.
(c)  One single   real root.}
\lb{fig0}
\end{figure}

With the approximate solutions for both scalar and tensor perturbations (Note that for both scalar and tensor modes, we have imposed the initial state as the Bunch-Davies vacuum, for details, see  \cite{Zhu}), one can compute the corresponding power spectra, which read,
\bqn\lb{pw}
\Delta^2(k)&\equiv &\frac{k^3}{2\pi}\left|\frac{\mu_k(y)}{z}\right|^2\nb\\
&=&\frac{k^2y}{4\pi^2z^2\nu}\mathcal{A}\exp \left(2\int_{y}^{y_0}\sqrt{g(\tilde{y})}d\tilde{y}\right),
\eqn
where the  factor $\mathcal{A}$ represents the trans-Planckian effects from the nonlinear terms in dispersion relation (\ref{omega}), could be amplified by the non-adiabatic evolution of inflationary perturbations, and is given by
\bqn
\mathcal{A} &\equiv& (1+2e^{\pi\xi_{0}^{2}}+2e^{\pi \xi_0^{2}/2}\sqrt{1+e^{\pi\xi_0^{2}}}\cos{2\mathfrak{B}}),
\eqn
with
\bqn
\xi_0^2&\equiv&\frac{2}{\pi}\int_{y_1}^{y_2} \sqrt{g(y)}dy,\\
\mathfrak{B}&\equiv& \int_{y_0}^{\text{Re}(y_1)} \sqrt{-g(y)}dy+\phi( \xi_0^2/2),
\eqn
where $\phi(x)\equiv x/2-(x/4)\ln{x^2}+\text{ph}\Gamma(ix+1/2)/2$ with $\text{ph}\Gamma(ix+1/2)$ is the phase of the Gamma function $\Gamma(ix+1/2)$, which is zero when $x=0$, and is determined by continuity otherwise. In the above the quantity $\xi_0^2$  depends on the properties of the turning points. $\xi_0^2$ could be positive, zero, and negative, corresponding to the facts that $g(y)$ has three real turning points, one single turning point and one double turning point, and one single and two complex turning points, respectively. It is obviously that when $\xi_0^2$ is positively large, the power spectra is amplified due to the non-adiabatic evolution, and when $\xi_0^2$ is negatively large, it means the adiabatic condition for the equation of motion is satisfied, and for this case the modified factor $\mathcal{A}$ is order of  $1$.

In order to study the effects of the nonlinear terms appearing in dispersion relations, represented by the $b_1$ and $b_2$ terms, one needs to perform the integral in Eq.(\ref{pw}) explicitly. For the scalar modes we identify the curvature fluctuation as $\mathcal{R}^2=\mu^2_k/z^2_s$ with $z_s=\sqrt{2\varepsilon} a$, and $h^2=8\mu^2_k/z_t^2$ for the tensor modes with $z_t=a$. Thus the general expression of tensor-to-scalar ratio can be written as
\bqn
r&=&\frac{8\Delta_t^2(k)}{\Delta^2_s(k)}=r_{\text{GR}} \left(\frac{\mathcal{A}_t}{\mathcal{A}_s} \sigma_k\right),
\eqn
where $r_{\text{GR}}=16 \varepsilon$ represents the tensor-to-scalar ratio predicted in slow-roll inflation models with a linear dispersion relation $\omega^2_k = k^2$ obtained in general relativity, and
\bqn\lb{sigma-integral}
\sigma_k\equiv \exp\left(2\int_y^{y_0^t} \sqrt{g_t(x)}dx-2\int_y^{y_0^s} \sqrt{g_s(x)}dx\right),
\eqn
where $y_0^t,\;g_t(y)$ are associated quantities for tensor modes, while $y_0^s,\;g_s(y)$ for scalar modes \footnote{It should be noted that the constants $\hat b_1$ and $\hat b_2$ appearing in the dispersion relation (\ref{omega}) are independent and different for scalar and tensor perturbations, as shown in \cite{WM,KUY,Soda2009prl,Wang2010PRD,Huang2013,Zhu2013}.}. Comparing $r$ with that in GR, one see that the factor $ \mathcal{A}_t\sigma_k /\mathcal{A}_s$ represents the trans-Planckian effects on the tensor-to-scale ratio. To reconcile  the tension between BICEP2 and Planck, it is required that
\bqn\lb{rbound}
\frac{\mathcal{A}_t}{\mathcal{A}_s} \sigma_k \gtrsim {\cal{O}}(2).
\eqn  
In the following we will show that this condition can be easily achieved by properly choosing the parameters of the nonlinear terms in the dispersion relations. 

Let us first take a look at Eq.(\ref{rbound}), from which we can see that the above condition  can be realized by two possible ways. The first mechanism is to raise the factor $\mathcal{A}_t/\mathcal{A}_s$ by incorporating the non-adiabatic effects, while assuming $\sigma_k \simeq 1$. Note that the later condition can be easily realized by assuming that $\epsilon_*^2$ is very small.  
However, as discussed in  \cite{Collins,ADSS,back-reaction}, once the non-adiabatic evolution is involved, a curial question is whether the back-reaction of the non-adiabatic modes is small enough to allow inflation to proceed. According to the analysis given in  \cite{back-reaction}, one has to constrain the ratio of the modification of scalar and tensor power spectra to the limits,
\bqn\lb{bound}
0.69^2<\frac{|\alpha+\beta|^2}{|\tilde\alpha+\tilde \beta|^2}<1.44^2,
\eqn
in order for the inflation to last for an enough long time, where $\beta$ and $\tilde \beta$ represent, respectively,  the Bogoliubov coefficients of the scalar and tensor modes, generated by the non-adiabatic evolution of the inflation. Here it is easy to identify that $\mathcal{A}_t =|\alpha+\beta|^2$ and $\mathcal{A}_s=|\tilde \alpha+\tilde \beta|^2$, thus one requires 
\bqn
0.48<\frac{\mathcal{A}_t}{\mathcal{A}_s}<2.07.
\eqn
Clearly, for ${\mathcal{A}_t}/{\mathcal{A}_s} \simeq {\cal{O}}(2)$, the requirement (\ref{rbound}) is fulfilled.   Even though this provides a possible way to reconcile the tension between Planck and
BICEP2, it should be noted that  the back-reaction of the non-adiabatic evolution is very difficult to calculate, and often depends on the specific theories,  tedious and  complicated analysis.

The second possibility is to raise the factor $\sigma_k$ by properly choosing the parameters of nonlinear terms in the dispersion relations. For the sake of the simplification, in this case let us assume that the adiabatic condition is satisfied during inflation for both the scalar and tensor modes, thus we have $\mathcal{A}_s \simeq 1 \simeq \mathcal{A}_t$. In order to evaluate  $\sigma_k$, one has to perform the integrals in Eq.(\ref{sigma-integral}) explicitly. However, these integrals, which involve nonlinear dispersion relations, are very difficult to compute. In order to see the effects of the nonlinear terms clearly, we can make some additional assumptions in the integrals. Let us first assume that the energy scale of inflation is less than the trans-Placnkian scale $M_*$, i.e., $\epsilon_*^2=H^2/M_*^2 < \mathcal{O}(1)$, thus the integrand of the integral in (\ref{sigma-integral}) can be expanded in terms of $\epsilon_*$. Up to order of $\mathcal{O}(\epsilon^4_*)$ and after some simple but tedious calculations we find that   
\bqn
\sigma_k&\simeq& \left(\frac{y_0^{t}}{\nu_t}\right)^{2\nu_t} \left(\frac{\nu_s}{y_0^s}\right)^{2\nu_s}\nb\\
&&\times \exp\left(-\frac{1}{3}b^t_1 \epsilon_*^2 y_0^{t3}-\frac{b_1^{t2}y_0^{t5}\epsilon_*^4}{5}+\frac{7b^t_2y_0^{t5}\epsilon_*^4}{15} \right)\nb\\
&&\times \exp\left(\frac{1}{3}b^s_1 \epsilon_*^2 y_0^{s3}+\frac{b_1^{s2}y_0^{s5}\epsilon_*^4}{5}-\frac{7b^s_2y_0^{s5}\epsilon_*^4}{15} \right),  ~~~~~
\eqn
with 
\bqn
y_0^t&\simeq& \nu_t+\frac{b_1^t\nu_t^3}{2} \epsilon_*^2+\frac{1}{8}(7b_1^{t2}-4b^t_2)\nu_t^5\epsilon_*^4+\mathcal{O}(\epsilon^6_*),\nb\\
y_0^s&\simeq& \nu_s+\frac{b^s_1\nu_s^3}{2} \epsilon_*^2+\frac{1}{8}(7b_1^{s2}-4b^s_2)\nu_s^5\epsilon_*^4+\mathcal{O}(\epsilon^6_*). 
\eqn
In the slow-roll approximations, we have $\nu_t\simeq \frac{3}{2}\simeq \nu_s$. Then,  up to the order of $\epsilon_*^2$ we find that
 \bqn\lb{sigma2}
 \sigma_k \simeq \left(\frac{1+9b_1^t\epsilon_*^2/8}{1+9b_1^s\epsilon_*^2/8}\right)^3\exp{\left[\frac{9}{8}(b_1^s - b_1^t)\epsilon^2_*\right]}.
 \eqn
In order to have   $\sigma_k \simeq {\cal{O}}(2)$, one can either raise the tensor spectrum or suppress the scalar spectrum. To raise the tensor spectrum, one needs to choose $b_1^t>0$ and $b_1^t\epsilon_*^2 \simeq \mathcal{O}(0.5)$. However, from Eq.(\ref{gy}) one sees that, when these   conditions are satisfied, $g(y)$ shall generically have three real turning points, which means that the adiabatic condition is violated and $\mathcal{A}_t$ shall   exceed the bound given in Eq.(\ref{bound}).  Another way to get $\sigma_k \simeq {\cal{O}}(2)$  is to   suppress the scalar spectrum. In this case, if one chooses $b_1^s < 0$ and $b_1^s \epsilon_*^2 \simeq
\mathcal{O} (0.5)$, $\sigma_k$ will raise from $1$ to $\mathcal{O}(2)$. For examples, from Eq.(\ref{sigma2}) we find that $\sigma_k \simeq 1.5$   for $9b_1^s/8\epsilon_*^2 \simeq -0.18$, and $\sigma_k \simeq 2$ 
  for $9b_1^s\epsilon_*^2/8 \simeq -0.28$, all with $b_1^t\epsilon_*^2\ll 1$. 
  
Here we would like to make some remarks about the suppressed scalar spectrum. First, to suppress the scalar spectrum, the parameter $\hat b_1$ is chosen to be positive, thus as mentioned in the above, the adiabatic condition is still satisfied. Consequently, the scale invariance of the spectrum is still robust and thus it is consistent with all the current observations. The only influence is the amplitude of the scalar spectrum. Second, as the $\hat b_1$ term only involves the fourth order spatial derivative terms in Ho\v{r}ava-Lifshitz theory, it is less constraint by the consistency analysis of instability, ghost, strong coupling, etc problems of the theory. Thus the parameter $\hat b_1$ chosen in the above does not lead any inconsistent issues. At last, we need to mentioned that the suppressed scalar spectrum have also been extensively discussed in refs.\cite{suppressed spectrum} by considering the quantum gravity effects.

 \section{Conclusions}
 
 In this brief report, we have taken the point of view that the  tensor-to-scalar ratio $r$ is big, as found by BICEP2, and  that the trans-Planckian effects become indeed important and need to be taken into account during the epoch of inflation, as indicated by the Lyth bound (\ref{Lyth}). Then, we have shown that,  even after  these effects are taken into account, almost scale-invariant perturbations can still be easily obtained \cite{WM,Wang2010PRD,Huang2013,Zhu2013}, and the inflation paradigm is robust. Moreover, these effects  also helps to relax the tension between the values of $r$ given, respectively,  by Planck    BICEP2, 
 whereby the  two problems mentioned in Introduction  are resolved.

\section*{Acknowledgements}
 This work is supported   in part by DOE  Grant, DE-FG02-10ER41692 (AW);
Ci\^encia Sem Fronteiras, No. 004/2013 - DRI/CAPES (AW);
NSFC No. 11375153 (AW), No. 11173021 (AW), No. 11047008 (TZ), No. 11105120 (TZ), 
and No. 11205133 (TZ).



\end{document}